\documentclass[reprint,amsmath,amssymb,aps,prl,floatfix,superscriptaddress]{revtex4-2}

\pdfmapfile{+cm.map}
\pdfmapfile{+symbols.map}
\pdfmapfile{+lm.map}
\usepackage[T1]{fontenc}
\usepackage{lmodern}
\usepackage{graphicx}
\usepackage{bm}
\usepackage{booktabs}
\usepackage{siunitx}
\usepackage{float}

\begin{document}

\title{Programmable Hard-X-Ray Spectral Comb Generation via Laser-Compton Scattering}

\author{Hao Zhang}
\email{haozh@g.ucla.edu}
\affiliation{Department of Electrical and Computer Engineering, University of California, Los Angeles, California 90095, USA}
\affiliation{SLAC National Accelerator Laboratory, Menlo Park, California 94025, USA}

\author{Ji Qiang}
\affiliation{Lawrence Berkeley National Laboratory, Berkeley, California 94720, USA}

\author{Shukui Zhang}
\affiliation{Thomas Jefferson National Accelerator Facility, Newport News, Virginia 23606, USA}

\author{Sergio Carbajo}

\affiliation{Department of Electrical and Computer Engineering, University of California, Los Angeles, California 90095, USA}
\affiliation{SLAC National Accelerator Laboratory, Menlo Park, California 94025, USA}
\affiliation{Physics and Astronomy Department, University of California, Los Angeles, CA 90095, USA}
\affiliation{California NanoSystems Institute, Los Angeles, CA 90095, USA}

\date{\today}

\begin{abstract}
Independent control of line spacing and offset underpins programmable frequency-comb spectroscopy, yet it has not been established for hard-x-ray sources based on inverse Compton scattering after electron transport and angular collection. Self-consistent multi-particle tracking and Li\'enard-Wiechert radiation calculations model a transported \SI{10}{\pico\coulomb} beam and its angle-integrated spectrum near \SI{11}{\kilo\electronvolt}. Laser pulse separation programs the comb spacing, while pulse-to-pulse phase translates the entire line grid at fixed spacing. The spacing model agrees within 0.52\%, while transport and optics change the phase-transfer slope by less than 0.2\%. At fixed electron spot size, the largest resolvable separation scales as the inverse square of normalized emittance; subnanometer bunches also produce a mode-locked train with \SI{13.9}{\atto\second} spacing. This two-coordinate programmability enables phase-stepped multicolor measurements and offers routes to hard-x-ray Ramsey interferometry, coherent control of core and nuclear excitations, and attosecond nonlinear spectroscopy.
\end{abstract}

\maketitle

Optical frequency combs use evenly spaced, phase-coherent lines for precision metrology \cite{udem2002optical,cundiff2003femtosecond}. For tooth index $q$, its angular frequency is $\omega_q=q\Delta\omega+\delta\omega$, where repetition controls the spacing $\Delta\omega$ and pulse-to-pulse phase controls the offset $\delta\omega$. High-harmonic generation has extended comb spectroscopy into the extreme ultraviolet \cite{gohle2005frequency,cingoz2012direct,benko2014extreme}. Mode-locked free-electron lasers (FELs), including a soft-x-ray demonstration and phase-locked \SI{9.7}{\kilo\electronvolt} pulse triplets, now approach hard-x-ray phase control \cite{thompson2008mode,hu2025demonstration,hu2025phlux,huang2026spectral}. Such pulse trains could drive Ramsey sequences and coherent wave packets at core-level or nuclear resonances \cite{lindroth2019challenges,shvydko2023scandium}. Inverse Compton scattering (ICS) Doppler-upshifts optical radiation from relativistic electrons into compact x-ray and gamma-ray beams \cite{esarey1993nonlinear,hartemann2005high,krafft2010compton,powers2014quasi,graves2014compact,sakai2024hard}. Optical pulse trains can imprint Compton or Thomson interference spectra, and delay scans control their spacing \cite{krajewska2014global,ilderton2020interference}. Laser chirp can compensate ponderomotive broadening, while carrier-envelope phase can alter nonlinear Thomson spectra \cite{ghebregziabher2013spectral,terzic2014narrow,seipt2015narrowband,rykovanov2016chirping,terzic2021chirping,ruijter2021cep}. Yet a programmable source also requires offset control at fixed spacing after realistic electron transport and finite-angle collection, together with a relation between beam phase space, comb density, and photon yield.

Two links are still needed for a programmable hard-X-ray source. First, the line offset must remain tunable at fixed spacing after electron transport and finite-angle collection, so that the spectral grid can move across a resonance without changing its sampling interval. Second, electron phase space and laser overlap must be connected to the resolvable comb density and photon yield. These controls enable a multiplexed scan: the spacing is matched to the energy intervals between several resonances, and stepping the offset sweeps the teeth across all of them simultaneously. A single exposure then records all channels under the same sample conditions, which is useful for transient states that evolve between shots or are altered by the probe.

\begin{figure}[t]
 \centering
 \includegraphics[width=0.98\linewidth]{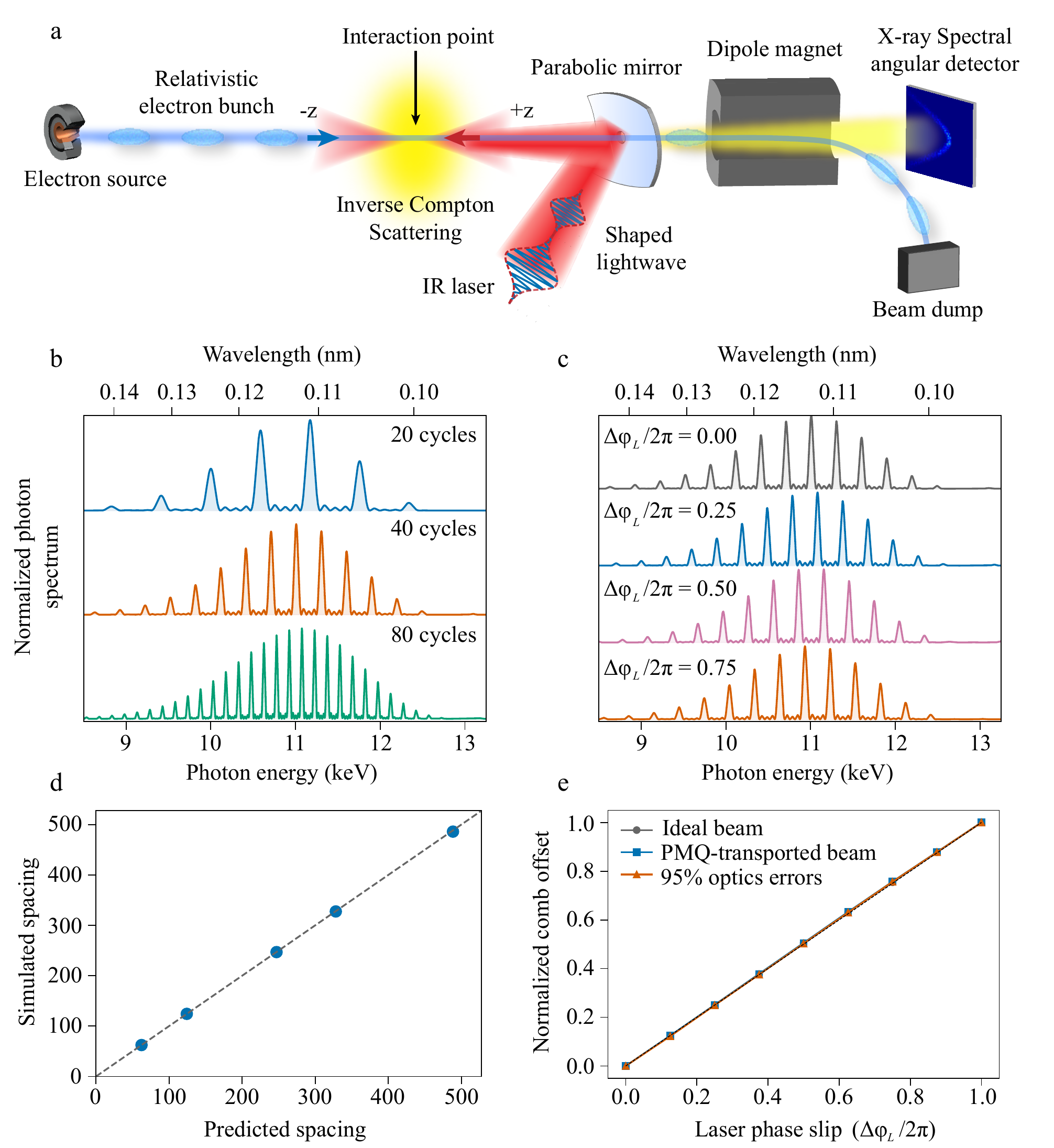}
 \caption{Programming an inverse-Compton comb. (a) Head-on collision and angle-resolved detection. (b) On-axis spectra for three pulse separations. (c) Phase-translated spectra at $N_{\mathrm{sep}}=40$. Curves are normalized separately and offset. (d) Spacing compared with Eq.~(\ref{eq:spacing}). (e) Normalized offset for ideal, transported, and optics-error beams.}
 \label{fig:programming}
\end{figure}

In this Letter, ImpactX supplies self-consistently transported electron distributions and a frequency-domain solver calculates their Li\'enard-Wiechert radiation \cite{huebl2022impactx,kan2023ics}. The working point uses a \SI{1030}{\nano\meter} laser, electron Lorentz factor $\gamma=50$, and normalized laser vector potential $a_0=0.55$. Scans of the subpulse separation $T_L$ and phase slip $\Delta\phi_L$ independently set the comb spacing and offset. Transport and optics alter the phase-transfer slope by less than 0.2\%, while divergence sets the largest resolvable separation. At an electron waist, the root-mean-square (rms) angular spread $\sigma_{x'}=\varepsilon_n/(\gamma\sigma_x)$ gives
\begin{equation*}
 N_{\mathrm{sep}}^{\max}\propto\frac{\sigma_x^2}{\varepsilon_n^2},\qquad
 N_\gamma\propto\frac{w_0^2}{\sigma_x^2},\qquad
 N_{\mathrm{sep}}^{\max}N_\gamma\propto\frac{w_0^2}{\varepsilon_n^2}.
\end{equation*}
Here $N_{\mathrm{sep}}^{\max}$ is the largest normalized pulse separation retaining spectral-comb contrast $\mathcal M\geq0.5$, $\sigma_x$ is the horizontal rms electron spot size, $w_0$ is the laser waist radius, $\varepsilon_n$ is the normalized transverse emittance, and $N_\gamma$ is the photon yield. The last two relations apply for $\sigma_x\gg w_0$.

Producing a periodic x-ray spectrum does not by itself make the source programmable. The spacing fixes the energy interval sampled by neighboring teeth, whereas the offset determines where that grid falls relative to a resonance. Spectroscopy requires both coordinates because a spacing scan also changes the number and placement of lines inside a finite bandwidth. An independent offset scan can interleave spectra without changing the instrumental resolution or the chosen channel separation. In an ICS source, this mapping must survive the correlated energy, angle, and position distributions of the electron bunch as well as integration over the detector aperture. Establishing that survival is the central goal of the calculations below.

Figure~\ref{fig:programming}(a) shows the collision geometry for a train of $M=5$ laser subpulses, while Figure~\ref{fig:programming}(b)-(e) summarize its programming laws. The incident laser field $E_L$ is
\begin{equation}
 E_L(u)=\sum_{m=0}^{M-1} A_m f(u-mT_L)
 \cos[\omega_L(u-mT_L)+m\Delta\phi_L],
 \label{eq:laser_train}
\end{equation}
where $u$ is retarded time; $m$ is the pulse index; $A_m$ is its amplitude; $f$ is the field envelope; $\omega_L$ is the laser angular frequency; $T_L$ is the pulse separation; and $\Delta\phi_L$ is the phase slip. The Gaussian envelope has an intensity full width at half maximum (FWHM) of \SI{10}{\femto\second}, or 2.91 laser cycles. Five pulses give resolved teeth; increasing $M$ mainly narrows them, as shown in the Supplemental Material. With laser period $T_0=2\pi/\omega_L$, normalized separation $N_{\mathrm{sep}}=T_L/T_0$, and effective envelope duration $\tau_{\mathrm{eff}}=3.10$ cycles, retardation gives the tooth spacing
\begin{equation}
 \begin{split}
 \frac{\Delta\omega_X}{\omega_L}&\simeq
 \frac{D_0}{N_{\mathrm{sep}}(1+\gamma^2\theta^2)
 +(a_0^2/2)\tau_{\mathrm{eff}}},\\
 D_0&=\frac{1+\beta}{1-\beta}\simeq4\gamma^2,
 \end{split}
 \label{eq:spacing}
\end{equation}
where $\Delta\omega_X$ is the x-ray tooth spacing, $\theta$ is the observation angle, $\beta$ is the electron speed in units of the speed of light $c$, and $D_0$ is the on-axis Doppler factor. The x-ray grid offset $\delta\omega_X$ follows
\begin{equation}
 \frac{\delta\omega_X}{\Delta\omega_X}
 \simeq\frac{\Delta\phi_L}{2\pi}\pmod{1}.
 \label{eq:offset}
\end{equation}
The retardation factor cancels from this ratio. The phase of the detrended spectral Fourier coefficient gives the grid displacement, and its normalized magnitude defines the spectral-comb contrast $\mathcal M$, from zero for no comb to unity for maximum modulation; derivations are given in End Matter and the Supplemental Material.

The two controls act on different parts of the interference condition. Changing $T_L$ changes the retarded arrival-time difference $\Delta t_X$ between adjacent scattered wave packets and therefore changes $\Delta\omega_X=2\pi/\Delta t_X$. Changing $\Delta\phi_L$ instead adds a phase to that condition and translates every maximum by the same fraction of one tooth spacing. The Doppler factor, nonlinear delay, and observation angle modify the absolute energy scale, but they enter the spacing and phase-induced displacement through the same $\Delta t_X$. Their cancellation in Eq.~(\ref{eq:offset}) is the physical basis for independent two-coordinate programming.

Figure~\ref{fig:programming}(b) and (d) show separations from 20 to 160 laser periods. The fitted spacings span $486\,\omega_L$ to $62.4\,\omega_L$ and agree with Eq.~(\ref{eq:spacing}) within 0.52\%. Duration and field-strength scans agree within 0.12\% and 0.093\%, respectively. A $\operatorname{sech}$ envelope matched in $\int f^2du$ agrees within 0.022\%, while an order-8 super-Gaussian differs by 2.6\% because its sharp edges fall outside the smooth-envelope approximation. The transported \SI{10}{\pico\coulomb} spectrum has an \SI{11}{\kilo\electronvolt} centroid within a \SI{2}{\milli\radian} collection half-angle. Additional checks appear in End Matter.

Figure~\ref{fig:programming}(c) shows that changing $\Delta\phi_L/2\pi$ from 0 to 0.75 translates the teeth at fixed spacing for $N_{\mathrm{sep}}=40$. Figure~\ref{fig:programming}(e) gives phase-transfer slopes of 1.003, 1.004, and 1.005 for the ideal, transported, and 95th-percentile optics-error beams, respectively. Extraction scans bound the transport- and optics-induced changes at 0.075\% and 0.172\% (Fig.~\ref{fig:end-phase-bias}).

Three electron distributions separate the sources of phase-transfer error. The ideal beam is generated at the interaction point with the nominal Gaussian moments and no transport correlations. The transported beam contains the full ImpactX output, while the optics-error case is the 95th-percentile permanent-magnet-quadrupole (PMQ) and entrance-Twiss sample near the accepted focus boundary. For each phase setting, a smooth spectral envelope is removed and the complex comb coefficient is evaluated at the fixed spacing. Unwrapping its phase and fitting the displacement against $\Delta\phi_L$ gives the transfer slope. Varying the smoothing width and analysis threshold produces absolute slopes from 0.989 to 1.006, which bounds sensitivity to the extraction procedure.

The phase-transfer slope remains stable because all electrons sample the phase slip imposed between neighboring laser pulses. Electron energy and angle shift the local carrier frequency and broaden the collected teeth, but the normalized displacement remains tied to the common pulse-train array factor. Transport and focusing errors therefore affect contrast and photon yield more strongly than the mapping between laser phase and x-ray line position. This separation allows the comb coordinates to be calibrated independently of later brightness optimization.

\begin{figure}[t]
 \centering
 \includegraphics[width=\linewidth]{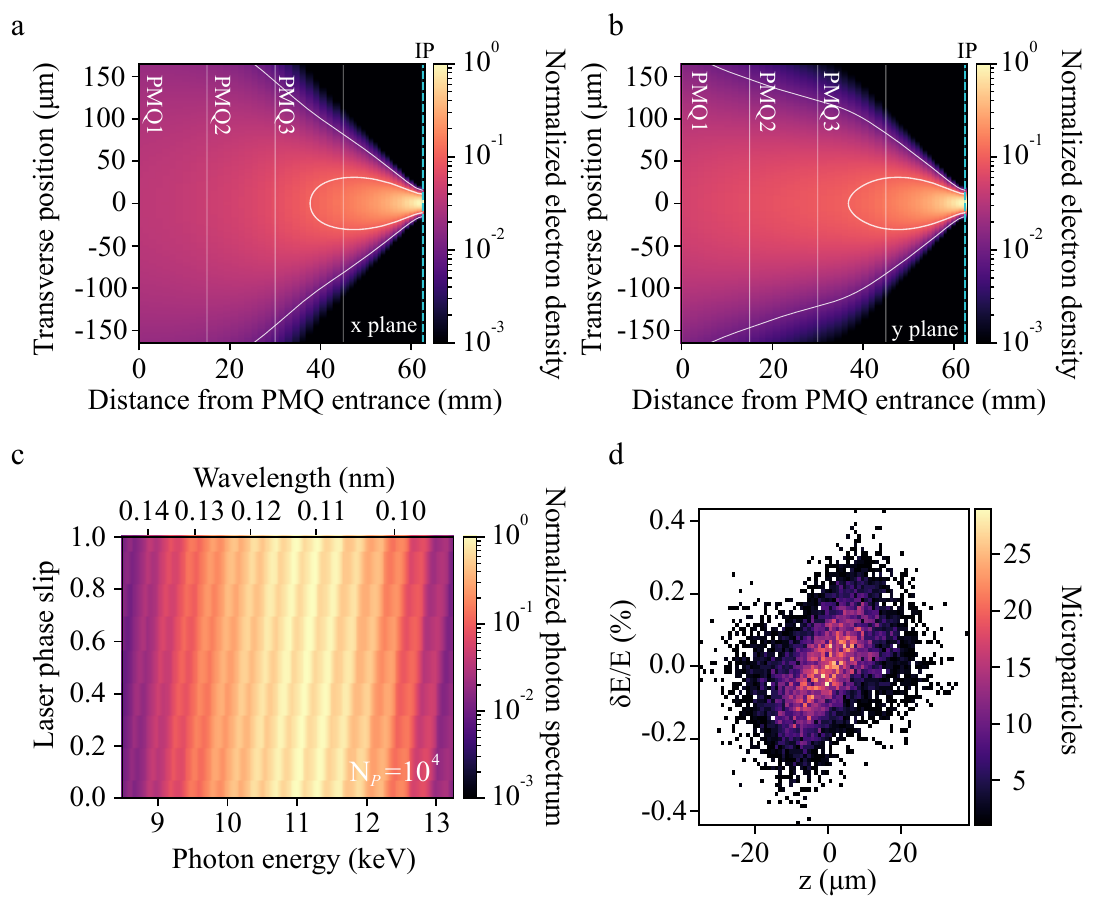}
 \caption{Electron transport and phase transfer. (a),(b) Transverse density through three PMQs and the final drift; white curves show rms envelopes and the dashed line marks the interaction point. (c) X-ray spectrum versus laser phase slip for the transported beam. (d) Longitudinal phase space used in the radiation calculation.}
 \label{fig:transport}
\end{figure}

Figure~\ref{fig:transport}(a) and (b) show the electron beam transport through a PMQ triplet in a self-consistent ImpactX simulation of a typical laser-Compton scattering study. The six-dimensional Gaussian input has \SI{10}{\pico\coulomb} charge, $10^4$ macroparticles, $\gamma=50$, and \SI{0.5}{\milli\meter\milli\radian} normalized emittance in each plane. Its rms bunch length is \SI{10}{\micro\meter}, and its rms relative energy spread is $10^{-3}$. The horizontal and vertical rms sizes are 146 and \SI{99.3}{\micro\meter}, with entrance Twiss parameters $\alpha_x=18.7$ and $\alpha_y=21.2$. Three \SI{15}{\milli\meter} soft-edge PMQs with gradients 84.1, 49.4, and \SI{-190.6}{\tesla\per\meter} precede a \SI{17.5}{\milli\meter} drift. A $64^3$ space-charge mesh with second-order particle shapes gives \SI{5.11}{\micro\meter} and \SI{5.10}{\micro\meter} rms focal sizes and \SI{1.96}{\milli\radian} and \SI{2.00}{\milli\radian} divergences. Figure~\ref{fig:transport}(c) shows the resulting phase-dependent x-ray spectrum, and Figure 2 (d) shows the longitudinal phase space supplied to the radiation calculation. Passing this distribution to the radiation calculation increases $\mathcal M$ by 1.62\% and lowers the on-axis yield by 1.28\% relative to the ideal beam. Convergence tests appear in the Supplemental Material.

\begin{figure}[t]
 \centering
 \includegraphics[width=\linewidth]{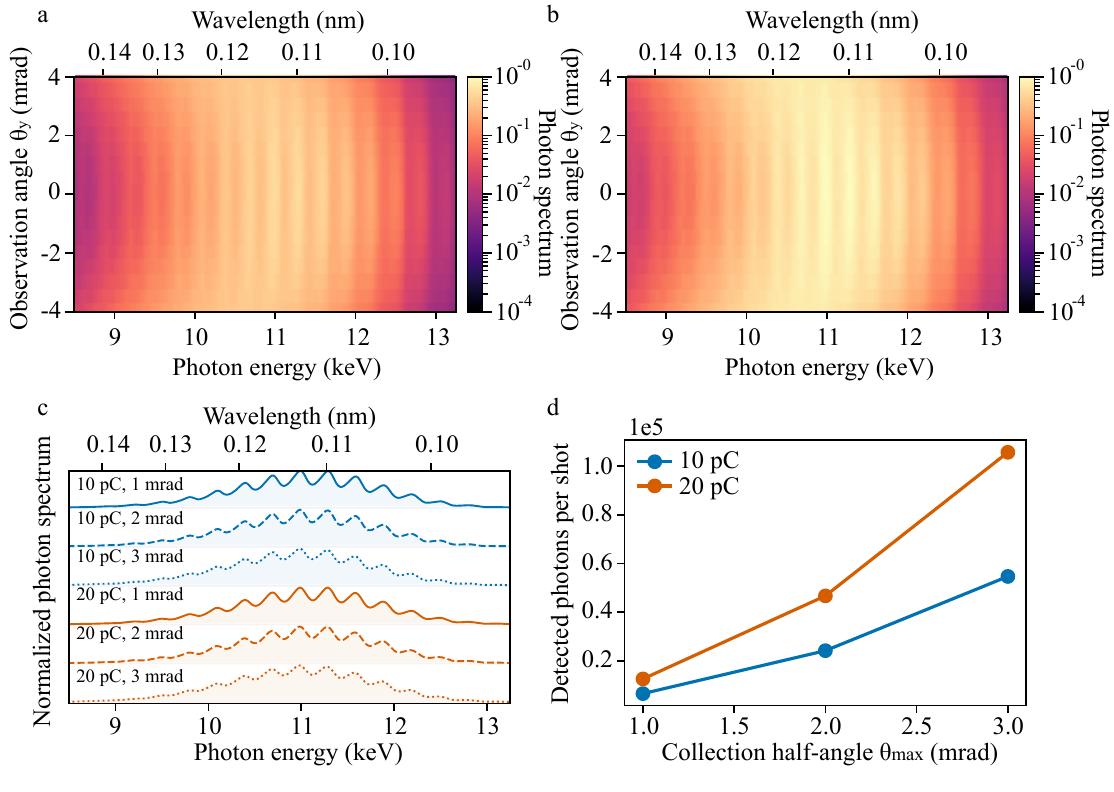}
 \caption{Angle-resolved spectra and yield. (a),(b) Photon distributions for 10 and \SI{20}{\pico\coulomb}. (c) Spectra integrated within 1, 2, and \SI{3}{\milli\radian} half-angles, normalized separately and offset. (d) Photons per shot versus half-angle; at \SI{2}{\milli\radian}, the 20-pC yield is 1.935 times the 10-pC value.}
 \label{fig:detector}
\end{figure}

Figure~\ref{fig:detector}(a) and (b) show the angle-resolved radiation for 10 and \SI{20}{\pico\coulomb}, while Figure~\ref{fig:detector}(d) gives the collected photon number. Li\'enard-Wiechert fields are calculated along each electron trajectory and summed incoherently over weighted macroparticles \cite{kan2023ics}. The angle-dependent Doppler shift curves the comb. Within a \SI{2}{\milli\radian} half-angle, raising the bunch charge from 10 to \SI{20}{\pico\coulomb} multiplies the yield by 1.935, a 3.3\% decrease per unit charge caused by space-charge-driven spot growth.

Figure~\ref{fig:detector}(c) shows the flux-resolution tradeoff for 1, 2, and \SI{3}{\milli\radian} half-angles. At \SI{2}{\milli\radian}, a 512-particle scan gives $\mathcal M=0.610$ at $N_{\mathrm{sep}}=120$ and 0.324 at 160, hence $N_{\mathrm{sep}}^{\max}\simeq135$. The divergence estimate gives $(\gamma^2\sigma_{x'}^2)^{-1}=(\sigma_x/\varepsilon_n)^2=104$, and fixed-spot emittance scans give $N_{\mathrm{sep}}^{\max}\propto\varepsilon_n^{-1.94\pm0.28}$, consistent with the predicted exponent $-2$. The working point has $\sigma_x/w_0=0.511$; spot-size-independent $N_{\mathrm{sep}}^{\max}N_\gamma$ requires the wide-beam limit.

This scaling follows from phase mixing across the electron angular distribution. For a waist divergence $\sigma_{x'}=\varepsilon_n/(\gamma\sigma_x)$, the Doppler shift varies by order $\gamma^2\sigma_{x'}^2$ across the bunch. Adjacent teeth remain resolvable while the accumulated shift over $N_{\mathrm{sep}}$ stays below one spacing, giving $N_{\mathrm{sep}}^{\max}\sim(\gamma^2\sigma_{x'}^2)^{-1}$. Increasing $\sigma_x$ therefore increases the accessible comb density, but reduces laser overlap and photon yield. In the wide-beam limit these effects cancel in their product, leaving emittance and laser waist as the useful design variables.

At $N_{\mathrm{sep}}=40$, the transported \SI{10}{\pico\coulomb} spectrum has $\mathcal M=0.709$, 16 resolved teeth, and $2.33\times10^4$ photons in the analyzed band. Ten bins per tooth spacing give a median of 121 photons per occupied bin and 9\% Poisson noise. For unit detection efficiency and zero background, two single-shot spectra give a three-standard-deviation phase-increment resolution of \SI{0.185}{\radian}. The required resolving power is 361, below the 1800 demonstrated from 6 to \SI{18}{\kilo\electronvolt} with a bent-quartz Cauchois spectrometer \cite{seely2012spectrometer}. Full-band acquisition requires a sufficient simultaneous energy window or spectral scanning.

The collection half-angle provides an experimental control complementary to the two laser coordinates. A smaller aperture suppresses angle-dependent Doppler broadening and preserves more densely spaced teeth, while a larger aperture collects more photons. Phase stepping cannot recover teeth erased by angular phase mixing, so the acceptance first sets the resolvable range of $N_{\mathrm{sep}}$. Within that range, $T_L$ selects the sampling interval and $\Delta\phi_L$ translates the grid. This ordering separates the photon-statistics requirement from the two spectroscopic programming operations.

\begin{figure}[t]
 \centering
 \includegraphics[width=\linewidth]{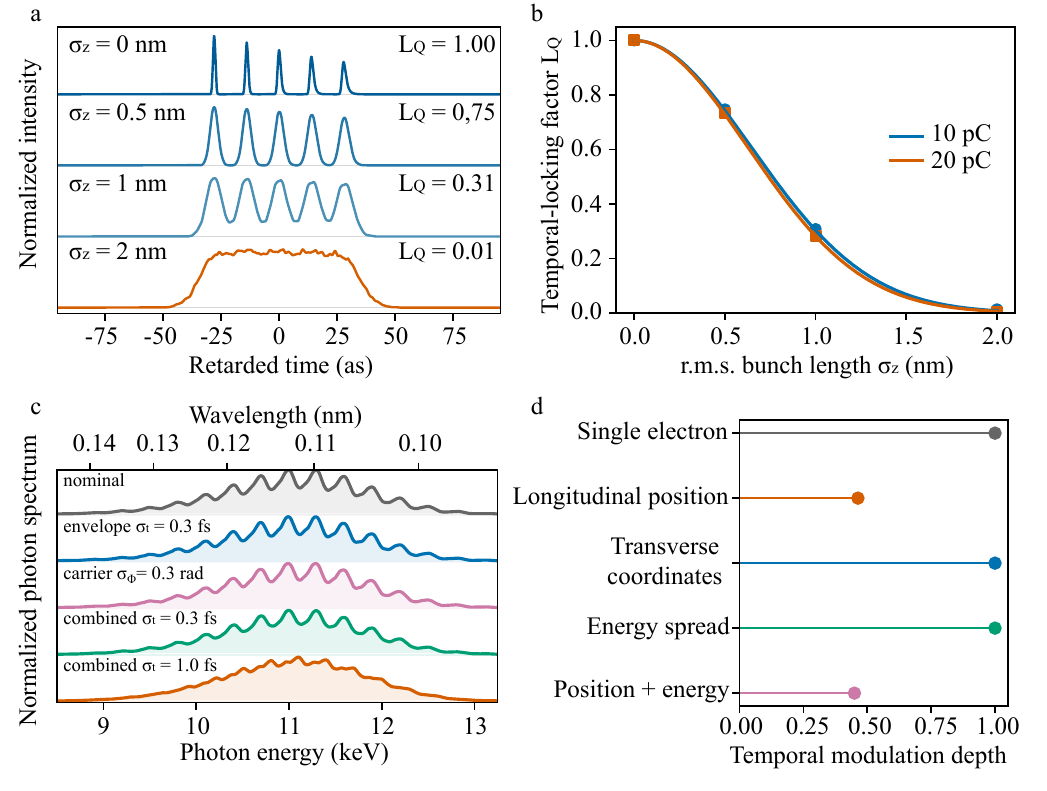}
 \caption{Temporal locking and tolerances. (a) Temporal intensity for a zero-length reference and 0.5, 1, and \SI{2}{\nano\meter} rms bunches. Curves are normalized separately and offset. (b) Locking factor for 10 and \SI{20}{\pico\coulomb}; curves fit Eq.~(\ref{eq:locking}). (c) Spectra under envelope, carrier, and physical-timing errors. (d) Modulation depth from individual electron coordinates.}
 \label{fig:locking}
\end{figure}

Figure~\ref{fig:locking}(a) shows the reconstructed temporal intensity, while Figure~\ref{fig:locking}(b) gives the locking factor that separates spectral modulation from temporal locking. After transforming each electron spectrum to retarded time, the normalized Fourier amplitude of the summed temporal intensity defines the charge-dependent locking factor $L_Q$, where the subscript $Q$ denotes bunch charge. For 10 and \SI{20}{\pico\coulomb}, $L_Q$ is 0.747 and 0.732 at \SI{0.5}{\nano\meter}, but 0.014 and 0.003 at \SI{2}{\nano\meter}. Its bunch-length dependence follows
\begin{equation}
 L_Q(\sigma_z)=\exp(-A_Q\sigma_z^2),
 \label{eq:locking}
\end{equation}
where $\sigma_z$ is the rms bunch length and $A_Q$ is the fitted locking coefficient at charge $Q$. Fits give $A_Q=1.19$ and $1.27\,\mathrm{nm}^{-2}$ for 10 and \SI{20}{\pico\coulomb}, close to the $1.14\,\mathrm{nm}^{-2}$ Gaussian estimate for \SI{13.9}{\atto\second} pulse spacing. The $L_Q=0.8$ boundaries are 0.433 and \SI{0.420}{\nano\meter}. The transported \SI{9.93}{\micro\meter} bunch retains $\mathcal M=0.709$ but has negligible temporal locking. Figure~\ref{fig:locking}(d) identifies longitudinal position as the main washout mechanism: it lowers the modulation depth from 1.00 to 0.463, or 0.450 when energy spread is included.

The small difference between the two fitted locking coefficients contains both charge-dependent transport and longitudinal correlations. A matched-parameter \SI{20}{\pico\coulomb} calculation gives $A_Q=1.25\,\mathrm{nm}^{-2}$ and accounts for 23\% of the difference from the 10-pC result. Figure~\ref{fig:locking}(c) separates envelope timing, carrier phase, and physical timing within one five-pulse train. At rms timing error $\sigma_t=\SI{0.3}{\femto\second}$, envelope timing retains a spectral correlation of $0.998\pm0.001$, whereas physical timing gives $0.96\pm0.03$ and introduces \SI{0.549}{\radian} rms carrier phase through $\omega_L\sigma_t$.

The distinction between spectral and temporal coherence is physical. A longitudinal shift of one electron translates its scattered pulse train in time without changing the train's spectral intensity. Incoherent addition can therefore preserve resolved spectral teeth even when the temporal peaks from different electrons no longer overlap. Temporal mode locking requires the electron-bunch length to remain below a fraction of the \SI{13.9}{\atto\second} x-ray period, whereas spectral programmability tolerates the much longer transported bunch.

The perturbations in Fig.~\ref{fig:locking}(c) represent distinct experimental errors. Envelope timing changes the pulse delays while compensating the carrier phase, carrier jitter changes phase at fixed envelope, and physical timing shifts both together. Because the optical carrier oscillates within the envelope, a subfemtosecond physical shift can produce a sizable phase error even when the pulse overlap changes little. Amplitude errors instead modify the relative weights of the five interfering contributions. The correlation $\rho$ records fine spectral structure and tooth displacement, $\mathcal M$ records the strength of the periodic modulation, and $N_\gamma$ records total yield. Their different responses show why stable photon number alone does not establish a reproducible programmable comb.

\begin{table}[t]
\caption{Within-train perturbations for $10^4$ particles. Here $\rho$ is correlation with the nominal detrended spectrum, $\Delta\mathcal M=100(\mathcal M/\mathcal M_0-1)$, and $\Delta N_\gamma=100(N_\gamma/N_{\gamma,0}-1)$; subscript 0 denotes the unperturbed train. Entries are means $\pm$ one standard deviation over 5 or 20 patterns.}
\label{tab:tolerance}
\setlength{\tabcolsep}{4pt}
\begin{ruledtabular}
\begin{tabular}{lccc}
Perturbation & $\rho$ & $\Delta\mathcal M$ (\%) & $\Delta N_\gamma$ (\%) \\
\hline
Envelope, \SI{0.3}{\femto\second} & $0.998\pm0.001$ & $-0.3\pm1.9$ & $+0.04\pm0.04$ \\
Carrier, \SI{0.3}{\radian} & $0.993\pm0.007$ & $-0.3\pm1.4$ & $+0.02\pm0.02$ \\
Amplitude, 10\% & $0.995\pm0.002$ & $-6.7\pm1.8$ & $+0.5\pm0.6$ \\
Physical, \SI{0.3}{\femto\second} & $0.957\pm0.032$ & $-3.1\pm2.5$ & $+0.001\pm0.002$ \\
Physical, \SI{1}{\femto\second} & $-0.16\pm0.32$ & $-18.9\pm22.1$ & $0.000\pm0.007$ \\
\end{tabular}
\end{ruledtabular}
\end{table}

Figure~\ref{fig:locking}(c) and Table~\ref{tab:tolerance} quantify pulse-train errors. A \SI{1}{\femto\second} physical-timing error lowers mean contrast by 18.9\% and decorrelates the teeth, while a 10\% amplitude error lowers contrast by 6.7\%. Yield changes remain within 0.5\%. At \SI{2}{\milli\radian} acceptance, the \SI{61}{\electronvolt} tooth width gives a spectral quality factor of 180; shot-to-shot phase stabilization and absolute frequency referencing remain experimental requirements.

The independent controls support a phase-stepped multicolor protocol. The pulse separation is first chosen so that the tooth spacing is commensurate with the energy intervals between selected absorption or scattering features. Phase steps then translate the full grid by known fractions of the spacing and fill the intervening energies. Ratios between teeth can provide simultaneous reference channels, while each exposure samples the same transient state during shock compression, a light-induced phase transition, or irreversible chemical change.

Temporal mode locking supports a second class of experiments. In the compressed-bunch regime, successive hard-X-ray bursts arrive every \SI{13.9}{\atto\second} with a programmed phase. Repeated interactions within a coherence time could drive Ramsey sequences on core-level or nuclear transitions. Phase cycling could prepare and read electronic or nuclear wave packets and separate quantum pathways in a nonlinear response. The pulse train also provides an attosecond stroboscope for charge migration, spin switching, and bond rearrangement, and could enable multidimensional hard-x-ray spectroscopy \cite{lindroth2019challenges,hu2025demonstration,hu2025phlux,shvydko2023scandium}. Longer laser trains would narrow the teeth, lower emittance would increase the resolvable density, and absolute frequency referencing would place the translated grid on a calibrated energy axis.

Mode-locked FELs create periodic spectra through gain modulation and structured electron-beam dynamics \cite{thompson2008mode,hu2025demonstration,huang2026spectral}. The present ICS scheme instead uses an externally prepared optical pulse train as its timing and phase reference. The laser sets the two comb coordinates before the collision, while the electron phase space determines tooth visibility and photon yield. Spacing and offset can therefore be reprogrammed without changing the PMQ lattice or electron energy. In the subnanometer regime, the same phase map connects the programmable frequency grid to a mode-locked temporal train.

A resonance experiment can use these controls hierarchically. The electron energy and laser wavelength place the overall ICS band, $T_L$ sets the interval between sampled energies, and phase steps translate the teeth across the target features. Near a narrow core-level or nuclear line, neighboring teeth can monitor the background or additional transitions during the same exposure. When the carrier is tuned near the \SI{12.4}{\kilo\electronvolt} $^{45}$Sc transition, the same protocol could map the nuclear response while preserving simultaneous reference channels \cite{shvydko2023scandium}. In the temporal-locking regime, repeated pulses add a controlled excitation history and supply Ramsey delays far shorter than those produced by mechanical x-ray optics. The same two laser variables thereby connect frequency-domain multiplexing and time-domain coherent control.

In conclusion, laser separation and phase independently control the spacing and offset of an inverse-Compton hard-x-ray comb after electron transport and angular collection. Electron divergence limits comb density, with $N_{\mathrm{sep}}^{\max}N_\gamma\propto w_0^2/\varepsilon_n^2$ in the wide-beam limit. Subnanometer electron bunches extend the source into temporal mode locking. Applications include transient multicolor measurements, hard-x-ray Ramsey interferometry, coherent control of core and nuclear excitations, and multidimensional hard-x-ray spectroscopy.

% \begin{acknowledgments}
% \textit{Acknowledgments}. Support was provided by the U.S. Department of Energy, Office of Science, Office of Basic Energy Sciences (DE-AC02-76SF00515, DE-AC02-05CH11231), Office of Nuclear Physics under contract DE-AC05-06OR23177, the National Science Foundation (2431903), and the Air Force Office of Scientific Research (FA9550-23-1-0409). The author thanks Yi-Kai Kan for discussions.
% \end{acknowledgments}

\begin{acknowledgments}
\textit{Acknowledgments}. Support was provided by the U.S. Department of Energy, Office of Science, Office of Basic Energy Sciences (DE-AC02-76SF00515), the National Science Foundation (2431903), and the Air Force Office of Scientific Research (FA9550-23-1-0409). The author thanks Yi-Kai Kan for discussions.
\end{acknowledgments}

\textit{Data availability}. The ImpactX and radiation-solver input decks, post-processing scripts, and data underlying the figures are available from the corresponding author upon reasonable request.

\bibliography{apssamp}

% \clearpage
\section*{End Matter}

\subsection*{Pulse-train spacing and offset laws}

Let $t$ be time, $z$ the longitudinal position, $c$ the speed of light, and $u=t+z/c$ the laser coordinate. The photon arrival time is $t_{\mathrm{obs}}=t-\hat{\mathbf n}\cdot\mathbf r/c$, up to a constant propagation delay, where $\hat{\mathbf n}$ is the observation direction and $\mathbf r$ is the electron position. Along the electron trajectory,
\begin{equation}
 \frac{dt_{\mathrm{obs}}}{du}
 =\frac{1-\hat{\mathbf n}\cdot\boldsymbol\beta}{1+\beta_z}
 \equiv\frac{1}{D(u,\theta)},
 \label{eq:end-time-map}
\end{equation}
where $\boldsymbol\beta$ is the electron velocity divided by $c$, $\beta_z$ is its longitudinal component, and $D(u,\theta)$ is the local Doppler factor:
\begin{equation}
 D(u,\theta)=\frac{1+\beta_z}{1-\hat{\mathbf n}\cdot\boldsymbol\beta}.
 \label{eq:end-doppler}
\end{equation}
For $\gamma\gg1$ and $\theta\ll1$, cycle averaging a linearly polarized plane wave with amplitude $a_0f(u)$ gives $\beta_z\simeq1-[1+a_0^2f^2(u)/2]/(2\gamma^2)$ and
\begin{equation}
 \frac{1}{D(u,\theta)}\simeq
 \frac{1+a_0^2f^2(u)/2+\gamma^2\theta^2}{4\gamma^2}.
 \label{eq:end-inverse-doppler}
\end{equation}
Corresponding points in adjacent subpulses are separated by $T_L=N_{\mathrm{sep}}T_0$ in $u$. Their observed delay is therefore
\begin{equation}
 \Delta t_X=\int_{u_0}^{u_0+T_L}\frac{du}{D(u,\theta)}.
 \label{eq:end-observed-delay}
\end{equation}
Here $u_0$ is the laser coordinate at a reference point in one subpulse. For identical, nonoverlapping smooth subpulses, define
\begin{equation}
 \tau_{\mathrm{eff}}=\frac{1}{T_0}\int_{-\infty}^{\infty}f^2(u)du.
 \label{eq:end-taueff}
\end{equation}
Integration of Eq.~(\ref{eq:end-inverse-doppler}) then gives
\begin{equation}
 \Delta t_X\simeq\frac{T_0}{D_0}\left[
 N_{\mathrm{sep}}(1+\gamma^2\theta^2)
 +\frac{a_0^2}{2}\tau_{\mathrm{eff}}\right],
 \qquad D_0=\frac{1+\beta}{1-\beta}.
 \label{eq:end-delay}
\end{equation}
Adjacent interference maxima are separated by $\Delta\omega_X=2\pi/\Delta t_X$, which yields Eq.~(\ref{eq:spacing}). If the field remains constant throughout the interval, $\tau_{\mathrm{eff}}=N_{\mathrm{sep}}$, and Eq.~(\ref{eq:end-delay}) recovers the usual Doppler denominator $1+a_0^2/2+\gamma^2\theta^2$. For isolated subpulses, the $a_0^2/2$ correction is instead weighted by the envelope integral.

At angular frequency $\omega$, the positive-frequency spectral field $\widetilde E_M(\omega)$ from $M$ identical subpulses contains the single-subpulse field $\widetilde E_1(\omega)$ and the array factor
\begin{equation}
 \widetilde E_M(\omega)=\widetilde E_1(\omega)
 \sum_{m=0}^{M-1}\exp\!\left[im(\omega\Delta t_X-\Delta\phi_L)\right].
 \label{eq:end-array-factor}
\end{equation}
For one electron, the scattered field contains a coherent contribution from every laser subpulse. The relative spectral phase between neighboring contributions has a frequency-dependent term set by $\Delta t_X$ and a constant term set by $\Delta\phi_L$. The first determines how rapidly the interference phase winds with photon energy; the second translates the maxima without changing that rate. The resulting five-pulse array factor converts this phase structure into resolved teeth whose width decreases as the train length increases. Because the single-pulse spectral envelope multiplies the array factor, changes to the laser strength or envelope can shift the carrier and redistribute intensity while preserving the two programming laws. This factorization explains why the laser controls remain separable when the absolute scattered frequency changes.

Its maxima satisfy $\omega_q\Delta t_X-\Delta\phi_L=2\pi q$. Hence
\begin{equation}
 \omega_{q+1}-\omega_q=\frac{2\pi}{\Delta t_X},
 \qquad
 \omega_q(\Delta\phi_L)-\omega_q(0)=\frac{\Delta\phi_L}{\Delta t_X},
 \label{eq:end-comb-coordinates}
\end{equation}
and their ratio gives Eq.~(\ref{eq:offset}). Thus the Doppler and envelope factors determine the tooth spacing but cancel from the normalized offset.

\subsection*{Transport and radiation workflow}

Self-consistent transport is essential because the radiation depends on the full six-dimensional distribution at the interaction point. Space charge and PMQ focusing jointly determine the transverse overlap with the laser, the angular spread entering the Doppler shift, and correlations between energy and longitudinal position. These quantities govern different observables: overlap controls photon yield, angular spread limits tooth resolution, and longitudinal phase space controls temporal locking. Passing the transported particles directly to the radiation solver keeps these effects correlated instead of replacing them with independent Gaussian broadening terms.

The frequency-domain radiation calculation evaluates the Li\'enard-Wiechert integral along each tracked trajectory. Contributions from the five laser subpulses are added at the field level for each electron, preserving the interference that forms its spectral comb. The weighted single-electron intensities are then summed incoherently over the macroparticles, consistent with a bunch that has no imposed x-ray-scale microbunching. This ordering retains the pulse-train interference while allowing energy spread, divergence, and transport correlations to broaden the ensemble spectrum. It also avoids artificial coherence between macroparticles when evaluating the photon yield. The spectra, angular maps, temporal reconstructions, and collected photon numbers therefore derive from the same transported phase-space distribution.

\subsection*{Spacing-law and Offset-extraction scan}

\begin{figure}[t]
 \centering
 \includegraphics[width=\columnwidth]{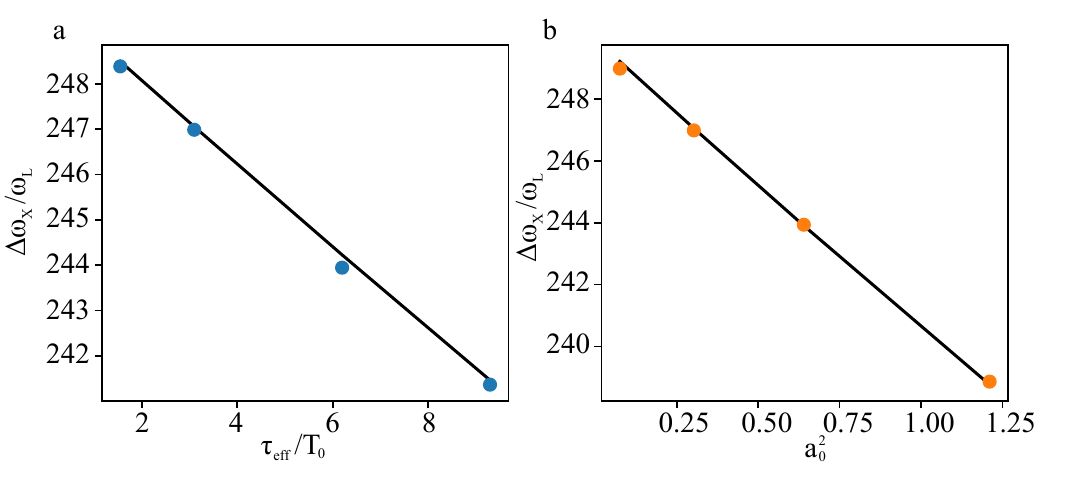}
 \caption{Independent validation of Eq.~(\ref{eq:spacing}) at $N_{\mathrm{sep}}=40$. Peak-fit spacing is shown versus effective duration at $a_0=0.55$ (a) and versus $a_0^2$ at an intensity FWHM of \SI{10}{\femto\second} (b). Solid curves use Eq.~(\ref{eq:spacing}) without fitted parameters.}
 \label{fig:end-spacing-tests}
\end{figure}

\begin{figure}[t]
 \centering
 \includegraphics[width=\columnwidth]{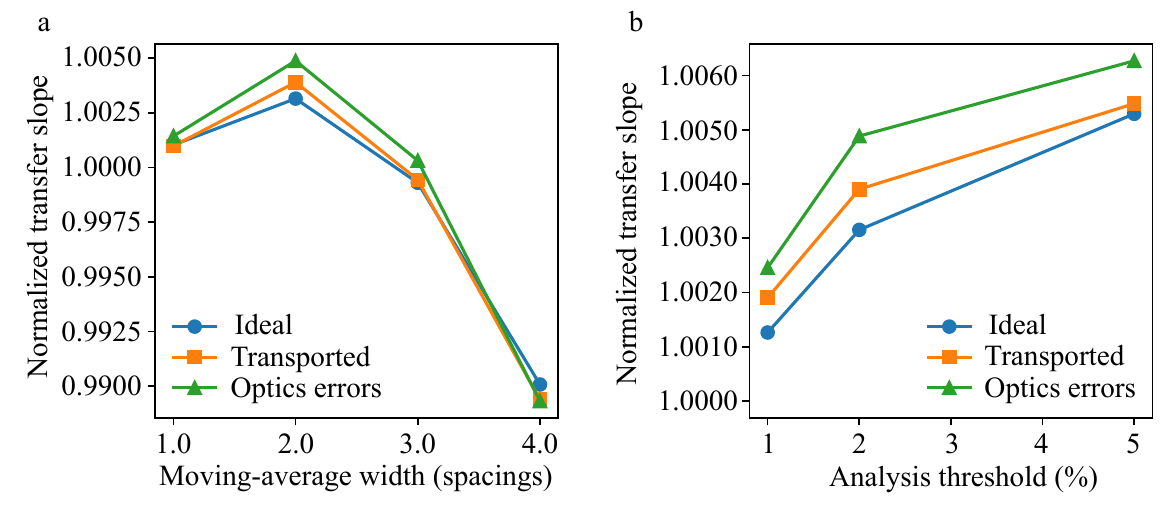}
 \caption{Normalized phase-transfer slope versus the moving-average width used to remove the spectral envelope (a) and the envelope threshold defining the analysis band (b). The spacing is fixed by Eq.~(\ref{eq:spacing}).}
 \label{fig:end-phase-bias}
\end{figure}

The Figure~\ref{fig:end-spacing-tests}(a-b) test the duration and field-strength dependences, respectively, of Eq.~(\ref{eq:spacing}). Table~\ref{tab:end-envelope} tests its envelope-shape dependence at fixed $\tau_{\mathrm{eff}}$. The left and right panels of Figure~\ref{fig:end-phase-bias} show the analysis variations underlying the absolute and relative slope ranges reported in Fig.~\ref{fig:programming}(e).

\begin{table}[t]
\caption{Envelope-shape validation at matched $\tau_{\mathrm{eff}}=3.098$ cycles, $a_0=0.55$, and $N_{\mathrm{sep}}=40$. Durations are intensity FWHM, and spacings are in units of $\omega_L$.}
\label{tab:end-envelope}
\begin{ruledtabular}
\begin{tabular}{cccc}
\centering
Envelope & Intensity FWHM (fs) & Eq.~(\ref{eq:spacing}) & Peak fit \\
\hline
Gaussian & 10 & 247.056 & 246.988 \\
$\operatorname{sech}$ & 9.382 & 247.056 & 247.042 \\
Super-Gaussian, order 8 & 10.797 & 247.056 & 240.65 \\
\end{tabular}
\end{ruledtabular}
\end{table}

\end{document}